
\input stromlo

\title A Theory of Changes and the Fundamental Plane

\shorttitle Changes and the Fundamental Plane

\author Stephen Levine

\shortauthor S. Levine

\affil Observatorio Astron\'omico Nacional, IA-UNAM, Ensenada, BC,
M\'exico.

\abstract Elliptical galaxies have been observed to cluster near a
ribbon along a two-dimensional plane (the Fundamental Plane of
Elliptical Galaxies) in the three dimensional space of effective
radius, effective surface brightness and central velocity dispersion.
This observed clustering holds for galaxies spanning several orders of
magnitude in size and total brightness and residing in a wide variety
of environments.  In order to understand the Fundamental Plane (FP),
it is important to know both why it should arise, and how it will
evolve.  This contribution addresses the second of these, how the FP
is likely to change with time.  In that context, we consider the
evolution due to galaxy--galaxy interactions, starting with arguments
based upon the virial theorem, and a discussion of the importance of
understanding the mass-to-light ratio as a function of position in the
galaxies.  The basic result of this argument is that most of any
global change in a galaxy due to a transformation (of whatever cause)
is mostly projected along the Fundamental Plane.  A secondary point is
that if the change includes a substantial change in the
mass--to--light ratio, a galaxy can move across the FP, which could
help explain the differences in observational properties between the
regular and dwarf galaxies.

\section Introduction

The clustering of elliptical galaxies along a two dimensional
manifold, the Fundamental Plane, in a 3-space of effective radius
($r_{\rm e}$), effective surface brightness ($\Sigma_{\rm e}$,
$\mu_{\rm e} = -2.5 \log \Sigma_{\rm e}$) and central velocity
dispersion ($\sigma_0$) (Djorgovski \& Davis 1987, Dressler
et~al. 1987, Bender, Burstein \& Faber 1992, 1993, Guzm{\'a}n, Lucey
\& Bower 1993, Pahre, 1996; see figure 1) requires an explanation.  An
additional point that has been noted before (e.g. Kormendy 1985,
Bender et~al. 1992,1993) is that the dwarf galaxies appear to behave
differently from the larger, regular elliptical galaxies.

\figureps[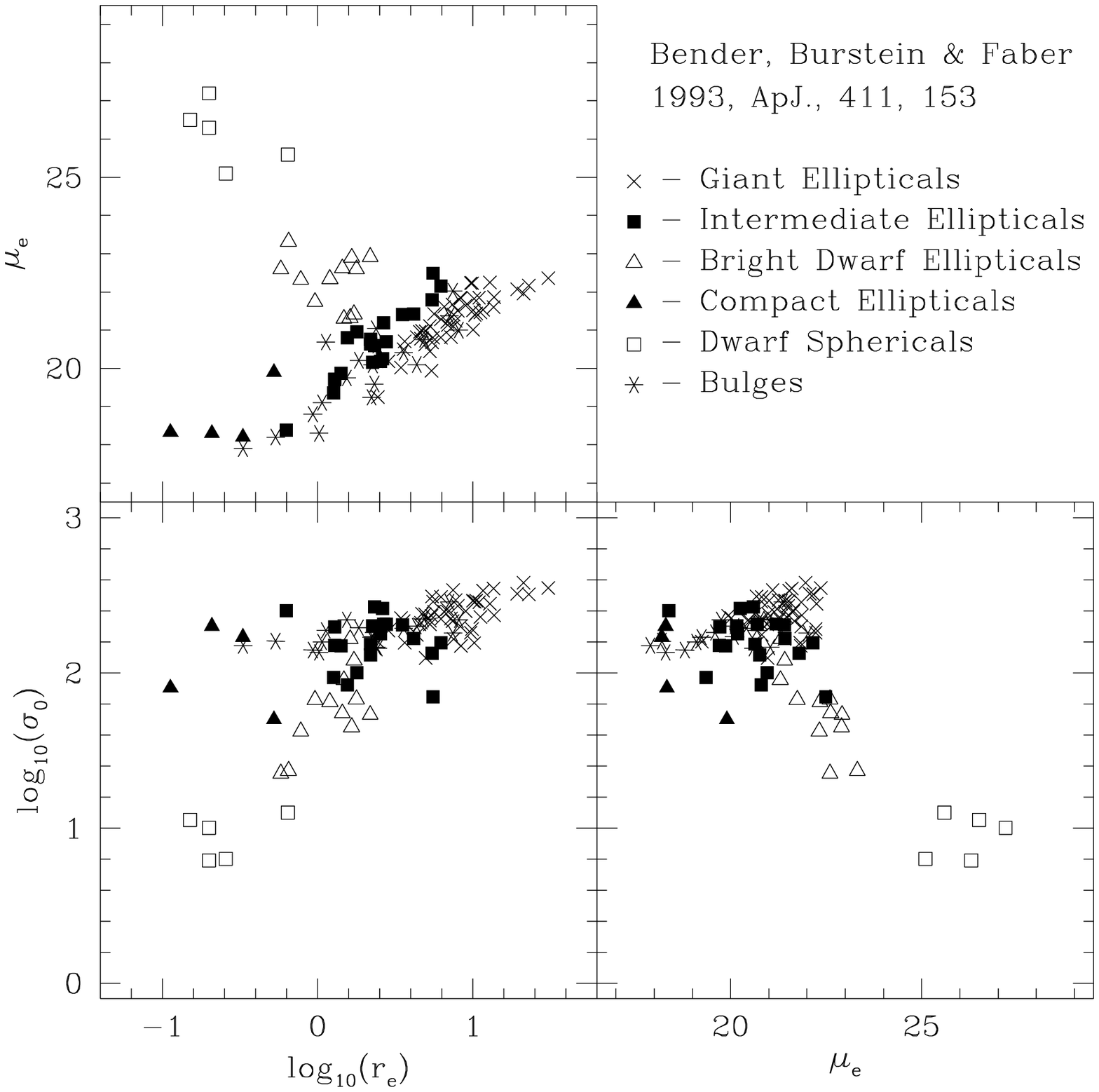,0.98\hsize] 1. Observed properties of a range of
elliptical systems (data are from table 1 of Bender, Burstein \& Faber
1993).  The dwarf galaxies ({\hfuzz=10pt \hbox to
0pt{$\sqcup$}$\sqcap$} \& $\triangle$) are thought to have formed
differently from the larger systems and/or may have evolved
differently, though through similar processes.

The fact that the galaxies lie on a plane has been explained by
assuming they are in virial equilibrium (e.g. Faber et al 1987, Bender
et al 1992), and, following the work of Poveda (1958), the distribution
of mass has been presumed to be directly proportional to the
distribution of light, so that a plane in mass becomes a plane in
light.  While a promising first approximation, this left open the
questions of why the galaxies appear to lie along a ``ribbon'' on the
manifold, why the dwarf galaxies appear discrepant, and why there was
some divergence from the plane defined by virial equilibrium.

This does not address the question of how a galaxy should appear to
move in this space as it evolves.  For example, Faber (1973) showed
how tidal stripping of the outer portions of an elliptical galaxy
could explain the existence of several elliptical galaxies with high
surface brightness and low luminosity.  In this case, assuming that
the outer portions of the galaxy simply vanish, the inner portion
comes to some new dynamical equilibrium, which can be characterized by
the virial theorem.  Levine \& Aguilar (1996) expanded upon the virial
theorem argument to allow for the possibility of a mass--to--light
ratio that is a function of position and showed how changes in the
underlying total mass and binding energy of a galaxy due to some
interaction can be related to the galaxy's observable properties.

In this contribution, I generalize the earlier argument of Levine \&
Aguilar (1996) and show how a better understanding of the fates of
luminous and dark matter as a result of interactions is crucial to
understanding the motions of galaxies in the space of observable
properties.  This can also help us to understand why ``regular''
elliptical galaxies exhibit a FP and why dwarf galaxies seem so at
odds with their larger cousins.  First I will discuss how to convert
light to mass, and then fold that into the argument of virial
equilibrium, and wrap up with a discussion of the projected
dispersions in observable properties.

\section{Light to Mass}

When we observe a galaxy, by definition, we are observing LIGHT.  If
we have taken spectra, then that light can be transformed into a
measure of the underlying potential of the galaxy through the velocity
dispersion.  We have no other direct measure of the underlying MASS of
the galaxy, unless we have some way to convert measurements of light
into mass.  The real dynamics is all based upon the distribution of
mass, including the assumption of virial equilibrium.  The simplest
and most often used assumption has been that the underlying mass is in
direct and constant proportion to the observed luminosity.  In this
case, the results predicated upon the dynamical argument of virial
equilibrium translate directly into observed light profiles.

Binney (1982b), Hjorth \& Madsen (1991, 1995) and Spergel \& Hernquist
(1992) have all noted that the natural product of an incomplete
violent relaxation process is an $r^{1/4}$ type profile.  Aguilar \&
White (1986) found empirically in their N-body simulations of
hyperbolic encounters that the radial density profile that resulted
from upsetting either a de~Vaucouleurs or a King model obeyed an
$r^{1/4}$ law.  Barnes and collaborators have found similar results in
merging encounters (see Barnes 1994 for an extended review of the
work).  These are arguments for why the underlying mass profile should
follow an $r^{1/4}$ law.  As Hjorth \& Madsen (1991) note in the very
beginning of their work, this is then consistent with the observed
profiles {\it if\/} the mass--to--light ratio is constant over the
galaxy.  This constancy is a sufficient, but not a necessary condition
for linking mass and light.

We are not limited to a constant mass--to--light ratio even if one
takes the observed robust result that the light in most elliptical
galaxies obeys an $r^{1/4}$ law, and the dynamical argument that the
mass should too.  In the more general case where mass and light each
satisfy possibly distinct de~Vaucouleurs laws, the ratio of surface
density [$\varsigma (r)$] to surface brightness [$\Sigma(r)$], $\alpha
(r) = {\varsigma (r) / \Sigma (r)} $ is an $r^{1/4}$ law function
$$
\alpha (r) = \alpha_{r = 0}
\exp {\left[-c \left( {r \over {\alpha_{\rm e}}}\right)^{1/4} \right]}
$$
where $c$ is a constant, and the ``effective M/L radius'' $
\alpha_{\rm e} $ and $\alpha_{r=0}$ are defined as 
$$
{1 \over {\alpha_{\rm e}^{1/4}}} \equiv 
\left[ {1 \over {r_{\rm g}^{1/4}}} - 
        {1 \over {r_{\rm e}^{1/4}}} \right] 
	~~{\hbox {\rm , \qquad and \qquad}}~~
\alpha_{r=0} = {{\varsigma(r=0)} \over {\Sigma(r=0)}} ~~. 
$$
$r_{\rm g}$ and $r_{\rm e}$ are the effective radii for the mass and
light profiles respectively.  In this manner, we can connect the
observed light with the underlying mass, and we end up with a more
flexible description of $\alpha (r)$.  Obviously if $r_{\rm g} =
r_{\rm e}$ then $\alpha$ is a constant and equal to the global $M/L$.
This is not to imply that this is necessarily the correct connection,
merely that the connection need not be a constant for theory and
observation to meet.

We need a better understanding of the variation of mass--to--light
ratio as a function of position.  With that, we can then construct a
more complete representation of the distribution of galaxy mass, check
the virial assumption and work towards a better understanding of how
mass--to--light {\it changes \/} over the course of galaxy--galaxy
interactions.

\section{Change in Mass to Change in Light}

In the case where an elliptical galaxy is changed under a
transformation that satisfies the following properties before and
after the change

\item{  (i)}{\it the galaxy is in virial equilibrium\/},
\item{ (ii)}{\it the mass and the light profiles have the same
2-parameter functional form\/},
\item{(iii)}{\it the functional forms of the radial surface
profiles do not change\/} and,
\item{ (iv)}{\it there exists a radius for which the local
mass--to--light ratio is constant over an encounter\/}\note{The
results given in Levine \& Aguilar (1996), are for the case when
the constant point is the center, $r=0$.}

\noindent
we can relate the changes in mass $M$, binding energy $E$, global
mass--to--light ratio $\langle \alpha \rangle$, and $\alpha_{r=0}$ to
changes in the observed galaxy properties.

The first criterion can be justified on theoretical grounds, since the
time scales for dynamical evolution for the inner regions of elliptical
galaxies are quite short compared to their lifetimes.  On
observational grounds the smooth distribution of light that is
prevalent in ellipticals argues for a least a quasi-equilibrium state
(see recent reviews by Bertin \& Stiavelli 1993, de~Zeeuw \& Franx
1991 and Binney 1982a).

The argument for a two parameter form (for the luminous profile) is
based upon the presumption that we can characterize an elliptical
galaxy solely by 2 parameters, e.g. total mass and internal binding
energy.  Djorgovski (1987) points out that the residuals in the
correlations of the various observed properties seen in elliptical
galaxies indicate that only two parameters are needed to describe
their global properties.  That the mass profile is of the same form is
an assumption based upon the arguments of the previous section and
clearly limited by our knowledge of the true variation of mass with
light.

In observations of clearly interacting spiral galaxies (e.g. Schweizer
1982), in the centers, an $r^{1/4}$ profile can be nicely fitted to
the radial surface brightness, indicating the robustness of the
$r^{1/4}$ profile as an outcome of strong interactions between
galaxies, even with non-elliptical progenitors.  This meshes well with
the previously noted theoretical reasons for expecting a similar
profile.  Assumption (iv) helps to restrict the solution space by a
dimension, and especially in the cases of moderate encounters, is
likely to be true near or at the center of the galaxy, since we would
expect only a quite strong transformation to have a large effect all
the way in as far as the center.

For a given variable $x$, we define the change in $x$ as $\Delta x
\equiv x_{\rm final} / x_{\rm initial}$.  The changes in the
observable galaxy properties are then related to the changes in the
global properties in the following manner:

$$
\eqalign{
\log(\Delta r_e) & = 2 \log(\Delta M) - \log(\Delta E) - 
	{1\over2}\log(\Delta\langle\alpha\rangle) +
	{1\over2}\log(\Delta\alpha_{r=0}) \cr
\log(\Delta \sigma_0) & = 
	{1\over2}\left[ -\log(\Delta M) + \log(\Delta E) \right] \cr
\log(\Delta \Sigma_e) & = -3\log(\Delta M) + 2\log(\Delta E) -
	\log(\Delta\langle\alpha\rangle) ~~.\cr
}
$$

\section{Interaction to Change in $M$, $E$, $\langle\alpha\rangle$ and
$\alpha_{r=0}$}

In order to be able to decipher the effects of tidal interactions upon
the observed properties of elliptical galaxies, we have undertaken a
series of numerical $N$-body simulations to determine empirically the
relationship between the parameters describing a tidal interaction and
the changes in global properties ($M$, $E$ etc) (Levine \& Aguilar
1997).  These are in essence interaction cross-sections.

To keep the project to a reasonable size, and still maintain decent
resolution in each simulation, we have limited ourselves to exploring
a range in galaxy mass ratio ($M_1/M_2 = 1,2,4,8$), impact velocity
($v_\infty = [\,^1\!/_2, 1, 2, 4]\times\sigma_0$) and impact parameter
($p_\infty = [0,1,2,4,8,16]\times r_{\rm g,1}$).  In all a set of 96
simulations have been run, covering both merging and hyperbolic
encounters.  Vergne \& Muzzio (1995) have also performed a large
series of simulations designed to explore the interaction parameter
space, but they lacked the necessary numerical resolution that we have
here.  Otherwise, almost all such work has been done using either only
equal mass galaxies or a rigid perturber or has only considered a very
few interactions (e.g. Aguilar \& White 1985, Capelato, de~Carvalho \&
Carlberg 1995).  The main thrust has often been to model a particular
interaction in more detail, rather than to map out the parameter space
(which admittedly is a very computationally expensive proposition).

These cross-sections can then be used for (among many other things) a
Monte-Carlo experiment similar to that used by Levine \& Aguilar
(1996), where they used an older set of equal mass interaction
cross-sections to try to compare the effect of tidal interactions with
the observed spread in galaxy properties.

\section{Discussion and Speculation}

Assuming that there is a way to convert interaction (or
transformation) parameters into changes in observable galaxy
properties, we can now ask, how does an ensemble of galaxies change
under interactions that satisfy the criteria given above?

For large galaxies subject to moderate non-merging encounters, we find
that the change in total mass and binding energy is fairly small.  In
addition, we expect that the change in central surface brightness and
central surface density will be small to non-existent
($\Delta\alpha_{r=0} \approx 1$).  Any differential change of luminous
versus dark matter will show up as a change in the overall
mass--to--light ratio.  The parallelograms in figure 2 show the
bounding box of change in observable parameters given a change in $M$
and $E$ equal to a factor of $\,^1\!/_2$.  The dashed line extending
out from the point (0,0) shows how the parallelogram would be
displaced if $\Delta\langle\alpha\rangle = \,^1\!/_2 {\rm \hbox{ and }
} 2$ and $\Delta\alpha_{r=0} = 1$.

\figureps[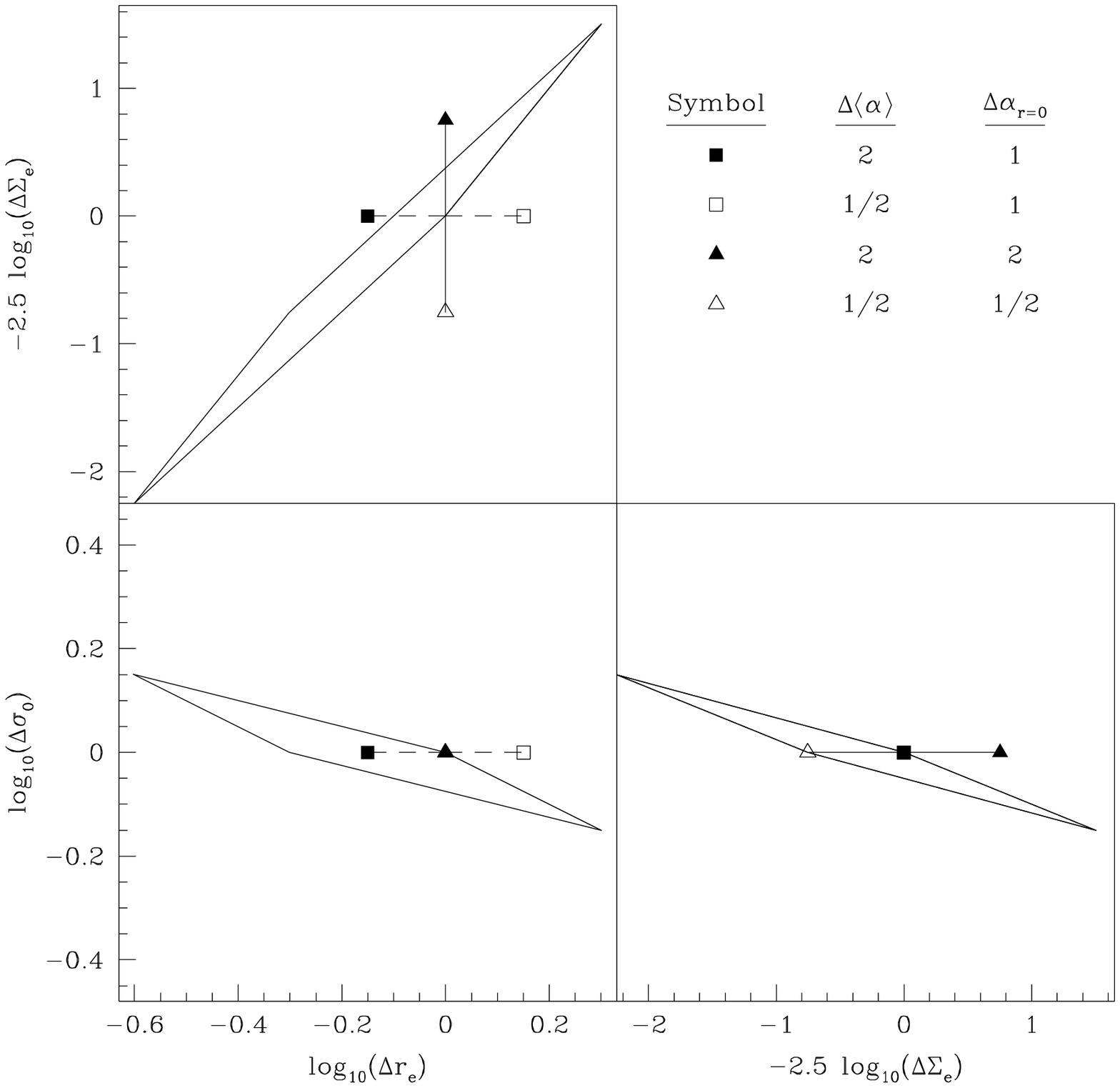,0.98\hsize] 2. The parallelogram which bounds
changes of up to $\Delta M = \,^1\!/_2$ and $\Delta E = \,^1\!/_2$ has
been projected into changes in the observable parameters.  The open
and filled square points show how the parallelogram would be displaced
if the global mass--to--light ratio were to change by factors of
$\,^1\!/_2$ and 2 respectively.  Similarly, the open and filled
triangles correspond to changes in BOTH $\Delta\langle\alpha\rangle$
and $\Delta\alpha_{r=0}$ of $\,^1\!/_2$ and 2.

For dwarf galaxies, the change in $M$ and $E$ can be quite
substantial.  Similarly, because the encounters can affect these
galaxies so dramatically, we suppose that any difference in the
efficiency of change with respect to light and dark mass will be felt
all the way into the center, and so $\Delta\alpha_{r=0} \approx
\Delta\langle\alpha\rangle$.  The solid line shows how the
parallelogram of change would be displaced if
$\Delta\langle\alpha\rangle = \Delta\alpha_{r=0} = \,^1\!/_2 {\rm
\hbox{ and }} 2$.  This is only a guide to the accessible regions of
the observable parameter space.  Based on the results of the simple
Monte-Carlo experiment of Levine \& Aguilar (1996), I expect that the
galaxies will fill only restricted parts of the available region.

I would venture to predict that the larger ellipticals will move in a
way that is close to parallel to the observed ribbon on the
fundamental plane, and that the dwarf ellipticals would move in a
direction similar to the dwarf ``branch'', which is not quite
orthogonal to the main ribbon (see figure 1).  In a manner analogous
to the ensemble of binary stars interacting with other stars, where
the hard binaries tend to get harder, and the soft binaries tend to
get softer, with a large enough sample of simulations, we may find
that the family of elliptical galaxies evolves, due to galaxy--galaxy
interaction, so that big galaxies get bigger, and smaller galaxies get
smaller.

\figureps[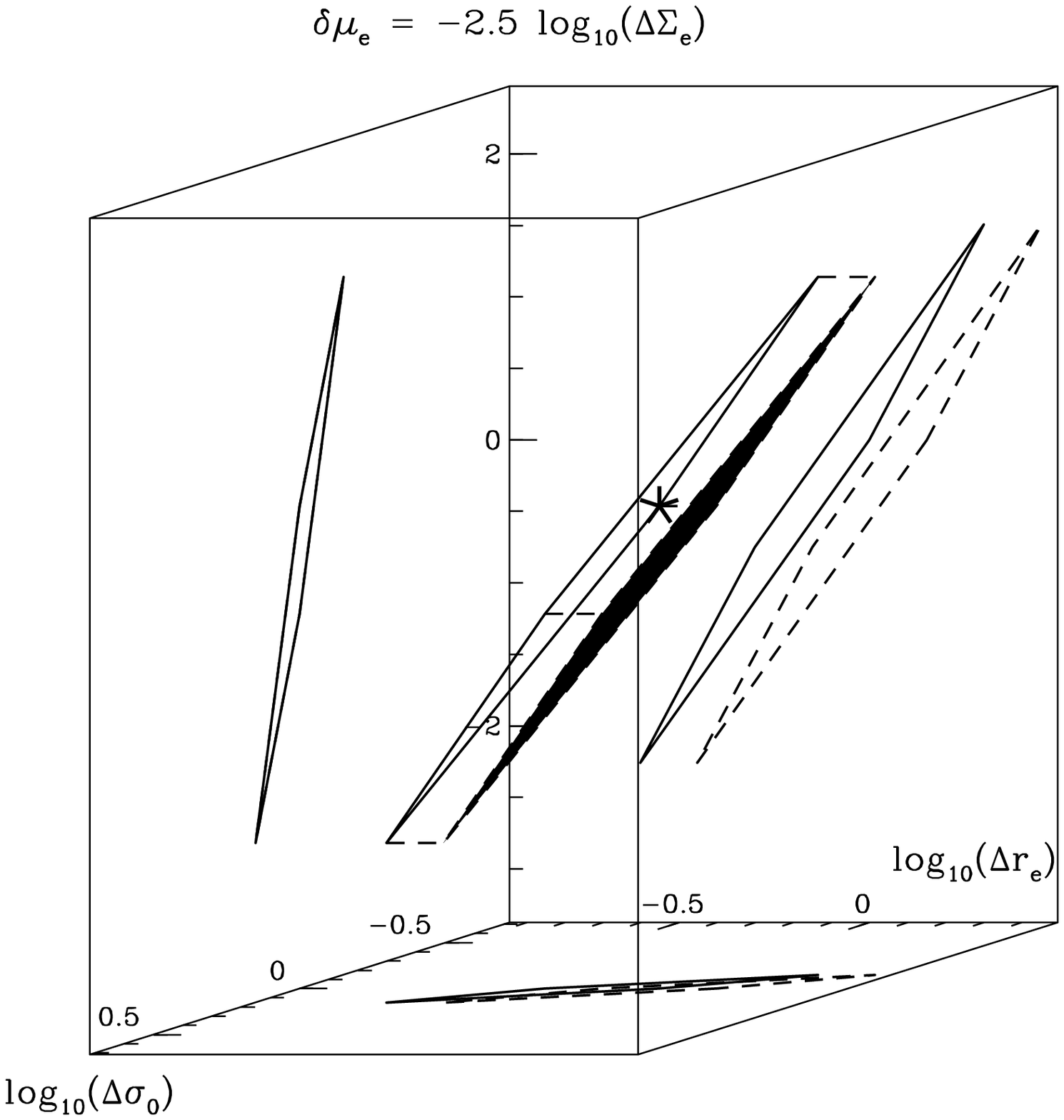,0.98\hsize] 3. The change parallelepiped
produced by a change of a factor of $\,^1\!/_2$ in $M$, $E$ and
$\langle\alpha\rangle$.  $\Delta\alpha_{r=0} = 1$. The five pointed
dot marks the initial point (0,0,0).  The face of the parallelepiped
where $\Delta\langle\alpha\rangle = 1$ is open and drawn in {\it
solid\/} lines, while that for $\Delta\langle\alpha\rangle = \,^1\!/_2
$ is shaded, and drawn in {\it dashed\/} lines.

The other aspect to note, is that once we allow the mass--to--light
ratio to change over a transformation, the galaxies as an ensemble are
no longer limited to an apparent plane, but can now fill a volume in
the space of {\it observable\/} parameters (a long, narrow volume, see
figure 3).  If we presume that there is some correlation between total
mass and total light, then perhaps the deviations from ``pure'' virial
equilibrium that people have noted are really a function of variation
in the evolution of the mass--to--light ratio and indicative of the
relative magnitude of changes effected upon galaxies during their
life.  The underlying mass might still be in equilibrium, but with a
variety of mass--to--light ratios the observed ensemble no longer
remains quite planar.

This does not explain how the galaxies reached the Fundamental Plane
in the first place, but it may help to show how the populations was
separated into the larger and smaller elliptical branches seen today,
and why the Fundamental Plane is not quite flat, but instead shows
some thickness.  Testing these speculations will depend upon (1) a
better understanding of how the mass--to--light ratio varies with
position in galaxies, (2) good interaction cross-sections for a wide
range of physical interaction parameters, and (3) an idea of what the
initial distribution of galaxies was in the observed parameter space.
The last of these is really a re-phrasing of the need to understand
how the Fundamental Plane formed in the first place.

\acknowl I am grateful to Luis Aguilar for interesting, amusing and
productive discussions.  This work has been supported by grant
\#3739--E from the Consejo Nacional de Ciencia y Tecnologia.

\references

{
\def\refindent{\goodbreak\par\noindent\parskip=1pt\hangindent=1pc\hangafter=1}

\refindent  Aguilar L. A., White S. D. M., 1985, ApJ, 295, 374

\refindent  Aguilar L. A., White S. D. M., 1986, ApJ, 307, 97

\refindent  Barnes J., 1994, in Mu\~noz-Tu\~n\'on C., S\'anchez F., eds., The
Formation and Evolution of Galaxies, Cambridge Univ. Press,
Cambridge, p.~399

\refindent  Bender R., Burstein D., Faber S., 1992, ApJ, 399, 462

\refindent  Bender R., Burstein D., Faber S., 1993, ApJ, 411, 153

\refindent  Bertin G., Stiavelli M., 1993, Rep. Prog. Phys., 56, 493

\refindent  Binney J., 1982a, ARAA, 20, 399

\refindent  Binney J., 1982b, MNRAS, 200, 951

\refindent  de~Zeeuw P. T., Franx M., 1991, ARAA, 29, 239

\refindent Capelato H. V., de~Carvalho R. R., Carlberg R. G., 1995,
ApJ, 451, 525

\refindent  Djorgovski S., 1987, in de Zeeuw P. T., ed., IAU Symp. \#127,
Structure and Dynamics of Elliptical Galaxies, Reidel, Dordrecht,
p. 79

\refindent  Djorgovski S., Davis M., 1987, ApJ, 313, 59

\refindent  Dressler A., Lynden-Bell D., Burstein D., Davies R., Faber S.,
Terlevich R., Wegner G., 1987, ApJ, 313, 42

\refindent  Faber S. M., 1973, ApJ, 179, 423

\refindent  Faber S. M., Dressler A., Davies R. L., Burstein D., Lynden-Bell
D., Terlevich R., Wegner G., 1987, in Faber S. M., ed., Nearly Normal
Galaxies, Springer-Verlag, NY, p. 175

\refindent  Guzm{\'a}n R., Lucey J., Bower R., 1993, in Danziger I. J.,
Zeilinger W. W., Kj{\"a}r K., eds, Structure, Dynamics and Chemical
Evolution of Elliptical Galaxies, ESO, Garching, p. 19

\refindent  Kormendy J., 1985, ApJ, 295, 73

\refindent  Levine S., Aguilar A., 1996, MNRAS, 280, L13

\refindent  Levine S., Aguilar A., 1997, in preparation

\refindent  Hjorth J., Madsen J., 1991, MNRAS, 253, 703

\refindent  Hjorth J., Madsen J., 1995, ApJ, 445, 55

\refindent  Pahre M., 1996, these proceedings

\refindent  Poveda A., 1958, Bull. Obs. Tonantzintla Tacubaya, 17, 3

\refindent  Schweizer F., 1982, ApJ, 252, 455

\refindent  Spergel D., Hernquist L., 1992, ApJ, 397, L75

\refindent  Vergne M., Muzzio J., 1995, MNRAS, 276, 439
}

\bye